\documentclass[letterpaper]{article} 
\usepackage{aaai25}  
\usepackage{times}  
\usepackage{helvet}  
\usepackage{courier}  
\usepackage[hyphens]{url}  
\usepackage{graphicx} 
\usepackage{natbib}  
\usepackage{caption} 
\usepackage{algorithm}
\usepackage{algorithmic}
\usepackage{amsmath}
\usepackage{newfloat}
\usepackage{listings}
\DeclareCaptionStyle{ruled}{labelfont=normalfont,labelsep=colon,strut=off} 
\floatstyle{ruled}
\newfloat{listing}{tb}{lst}{}
\floatname{listing}{Listing}
\title{AAAI Press Formatting Instructions \\for Authors Using \LaTeX{} --- A Guide}
\author{
    Written by AAAI Press Staff\textsuperscript{\rm 1}\thanks{With help from the AAAI Publications Committee.}\\
    AAAI Style Contributions by Pater Patel Schneider,
    Sunil Issar,\\
    J. Scott Penberthy,
    George Ferguson,
    Hans Guesgen,
    Francisco Cruz\equalcontrib,
    Marc Pujol-Gonzalez\equalcontrib
}

\title{Learning coupled
Allen-Cahn and Cahn-Hilliard phase-field equations using Physics-informed neural operator(PINO)}
\author {
    Gaijinliu Gangmei\textsuperscript{\rm 1,\rm 2},
    Santu Rana\textsuperscript{\rm 1},
    Bernard Rolfe\textsuperscript{\rm 1},
    Kishalay Mitra\textsuperscript{\rm 2},
    Saswata Bhattacharyya\textsuperscript{\rm 2}
}
\affiliations {
    \textsuperscript{\rm 1}Deakin University, Australia\\
    \textsuperscript{\rm 2}Indian Institute of Technology Hyderabad,India\\
}

\begin{document}

\maketitle

\begin{abstract}
Phase-field equations, mostly solved numerically, are known for capturing the mesoscale microstructural evolution of a material. However, such numerical solvers are computationally expensive as it needs to generate fine mesh systems to solve the complex Partial Differential Equations(PDEs) with good accuracy.  Therefore, we propose an alternative approach of predicting the microstructural evolution subjected to periodic boundary conditions using Physics informed Neural Operators (PINOs). 
 In this study, we have demonstrated the capability of PINO to predict the growth of  $\theta^{\prime}$ precipitates in Al-Cu  alloys by learning the operator as well as by solving three coupled physics equations simultaneously. The coupling is of two second-order Allen-Cahn equation and one fourth-order Cahn-Hilliard equation. We also found that using Fourier derivatives(pseudo-spectral method and Fourier extension) instead of Finite Difference Method improved  the Cahn-Hilliard equation loss by twelve orders of magnitude. Moreover, since differentiation is equivalent to multiplication in the Fourier domain, unlike Physics informed Neural Networks(PINNs), we can easily compute the fourth derivative of Cahn-Hilliard equation without converting it to coupled second order derivative. 
\end{abstract}


\section{Introduction}
The evolution of a microstructure can be described using complex PDEs such as Allen-Cahn and Cahn-Hilliard equation. These equations are traditionally solved using numerical methods such as pseudo-spectral method~\cite{liu2003phase}, finite element methods(FEM)~\cite{barrett1999finite}, and finite difference methods(FDM)~\cite{sun1995second}. However, solving complex multi-physics and high order PDEs using numerical methods are computationally expensive due to the requirement of fine mesh in both the spatial and time domain to obtain accurate solution. Because of the limitations involved in using numerical method for solving PDEs, recently researchers have started looking into Machine Learning(ML) approaches for solving PDEs. ML-based approaches for solvinge PDEs can be divided into two categories i.e., solution function approximation and operator learning. In operator learning the model learns the solution operator of a given family of parametric PDEs, e.g., Fourier Neural Operator(FNO)~\cite{li2020fourier}, Deep Operator Network(DeepONet)~\cite{lu2021learning}, and their  physics-informed counterparts~\cite{li2021physics, wang2021learning}. Whereas, in solution function approximation like PINNs~\cite{raissi2019physics}, the model parameterizes the solution function of a single instance of the PDE as a neural network. 
For each new instance of the PDEs, PINNs needs to train a new neural network~\cite{chen4761824pf}. This makes them impractical for applications where a PDE solution is needed for several parameter instances, such as those with varying shapes, initial or boundary conditions, coefficients, etc.~\cite{li2020fourier}. Such parametric PDEs can be learnt using FNO and DeepONet but both are data-driven, meaning that it requires a larger volume of training data to produce accurate results for complex problems. Generating the data is expensive as it involves repeated evaluation of experiments or high-fidelity numerical simulators. Therefore, by adding physics constraints and using some available data, PINO and physics-informed DeepONets learns the solution operator of a family of PDEs. It will make the model more robust with limited number of data and possibly improve its ability to capture non-linearities in the model evolution.

Physics-informed DeepONets learns the operator by using two neural networks known as the branch and the trunk network. Similar to PINNs, it uses automatic differentiation to calculate the derivatives of the physics which are memory intensive. Whereas, PINO is a variation of neural operators that learns the operator in the Fourier domain. The derivatives are also calculated in Fourier domain which simplifies the computation since differentiation is equivalent to multiplication in this domain. Therefore, in this study we have chosen PINO to learns the dynamics of growth of $\theta^{\prime}$ precipitates in Al-Cu alloy.  
The feasibility and computational advantage of PINOs have been proven previously in literature involving only first and second order PDEs~\cite{rosofsky2023applications,li2021physics}. Hence, in this work we determined the capability of PINO in solving three coupled PDEs: two Allen-Cahn and one Cahn-Hillard equations which are second-order and fourth-order PDEs respectively. The results from this study was compared with the high fidelity data obtained from computational phase-field method. 
We used PyTorch framework~\cite{paszke2019pytorch} to train the PINO model using two NVIDIA V100 32 GB GPU for 2D problem.

\section{Methodology}
\subsection{Overview}
\begin{figure*}[!h]
\centering
\includegraphics[width=0.8\textwidth]{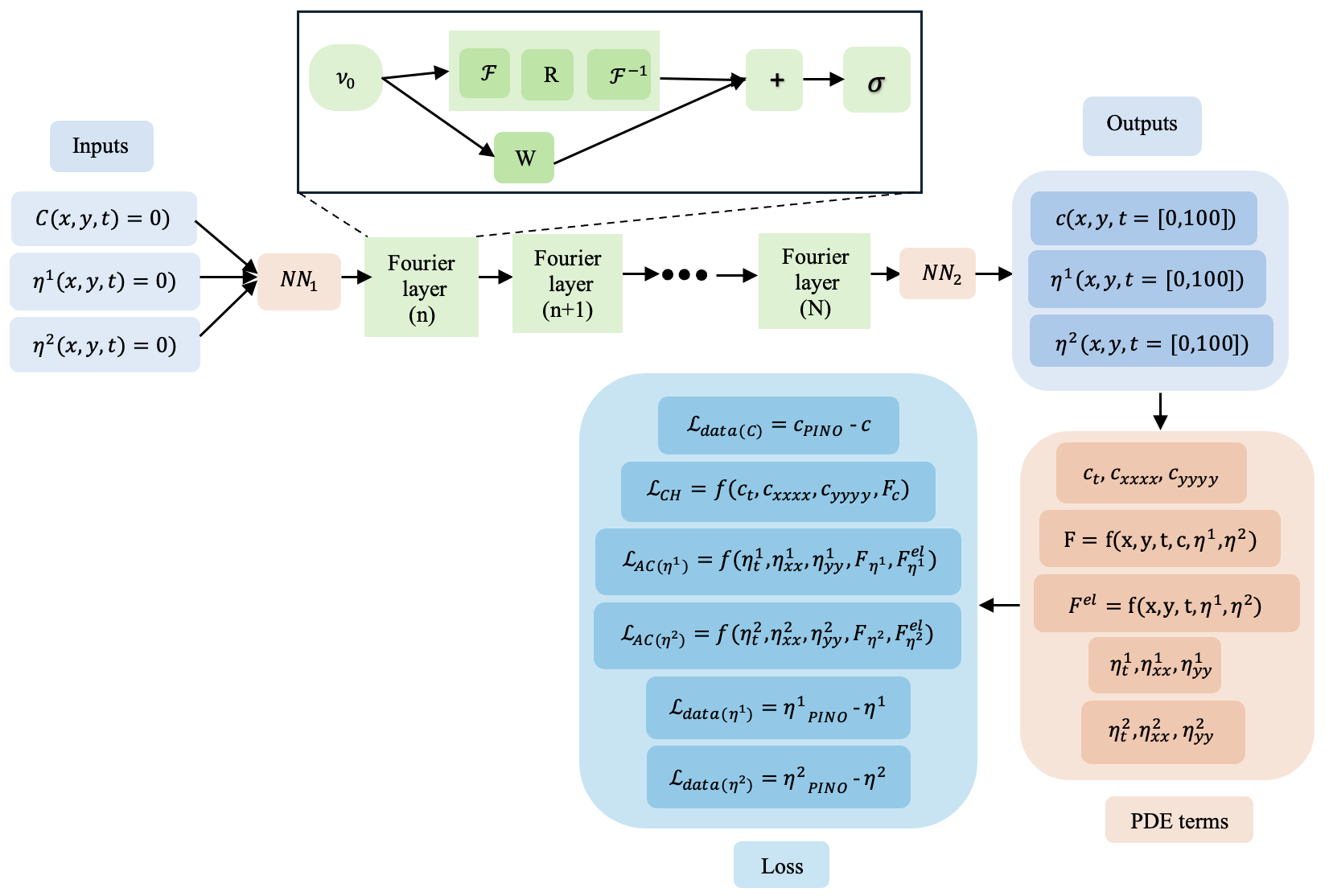}
\caption{Architecture of PINO for learning the dynamics of $\theta^{\prime}$ precipitation in Al-Cu alloys(2D).} \label{fig:PINO}
\end{figure*}

PINO uses the FNO architecture instead of feed forward neural network to overcome the challenges of PINNs. It learns the solution operator using the physics constraints as well as supervision from data. 
In this study, the growth of $\theta^{\prime}$ precipitates in a binary alloy is described by three phase-field variables namely, composition($c$) and two order parameter ($\eta_1$ and $\eta_2$) fields. Therefore, to learn the precipitate growth, PINO requires to have three inputs and corresponding three outputs at a given time. Furthermore, in the coupled physics problem, PINOs has to solve multiple equations simultaneously. The architecture of PINO for coupled fields is shown in Fig.\ref{fig:PINO}.

The neural operator of the PINO model has the form:
\begin{equation}
\mathcal{G}_{\theta}=NN_2\circ(W_N+\mathcal{K}_N)\circ\dots\circ\sigma(W_n+\mathcal{K}_n)\circ NN_1.
\end{equation}
Where, $NN_1$ is a feed-forward neural network that lifts the inputs to a higher dimension($\upsilon_0$), $W$ is the bias
term which keeps track of the non-periodic boundary conditions and $\mathcal{\kappa}$ is an integral kernel operator in frequency space which are multiplied with an activation function($\sigma$) at the  intermediate Fourier layer, and $NN_2$  is the feed-forward neural networks that projects back to the original dimension of the
solution ($c, \eta_1, \eta_2$) at the end. Once the solution from PINO model is obtained, loss function consisting of data loss and PDE loss  will be computed followed by gradient calculation and weight update.
The total loss function is defined as:
\begin{equation} \label{eq1}
\begin{split}
\mathcal{L}_{total}&=w_{data(\eta_1)}\mathcal{L}_{data(\eta_1)}+w_{pde(AC_{\eta_1})}\mathcal{L}_{pde(AC_{\eta_1})}\\& +w_{data(\eta_2)}\mathcal{L}_{data(\eta_2)}+w_{pde(AC_{\eta_2})}\mathcal{L}_{pde(AC_{\eta_2})}
\\&+w_{data(c)}\mathcal{L}_{data(c)}+w_{pde(CH)}\mathcal{L}_{pde(CH)}
\end{split}
\end{equation}
where $w$ are the weights for each loss terms. The loss term for initial
conditions is not added in the loss function as they are already embedded in the data we
are using to train the model. Moreover, for our precipitate growth problem with periodic boundary conditions we do not need to include a loss term for the boundary conditions(BCs) because the neural network architecture of FNO implements periodic BCs by default via fast Fourier transform (FFTs).

\section{Computational Experiments}
\subsection{Data generation}
Data with various combinations of supersaturated composition and seed were generated using a phase-field model~\cite{kumar2017enhancing} designed to investigate the evolution of Al-2Cu precipitates. In 2D representation, $\theta^{\prime}$ precipitates in Al-Cu alloy is defined by one composition field($c$) and two long-range order parameter  fields ($\eta_1$ and $\eta_2$). The growth of the precipitates is governed by a Cahn-Hilliard~\cite{cahn1961spinodal} equation and two Allen-Cahn~\cite{allen1979microscopic} equation as given in Eq.~\ref{eqn: CahnHilliard} and Eq. \ref{eqn: AllenCahn} respectively. In addition to chemical energy, elastic energy contribution to total energy is considered to get the desired $\theta^{\prime}$ morphology.
\begin{equation}\label{eqn: CahnHilliard}
\begin{split}
   \frac{\partial c}{\partial t}&=M\left[\frac{\partial^2}{\partial x^2}\left(\frac{\partial f}{\partial c}\right)+\frac{\partial^2}{\partial y^2}\left(\frac{\partial f}{\partial c}\right)\right]\\
   &-2\kappa_c M\left(\frac{\partial^4c}{\partial x^4}+\frac{\partial^4c}{\partial y^4}\right),
   \end{split}
\end{equation}
\begin{equation}\label{eqn: AllenCahn}
   \frac{\partial \eta_i}{\partial t}=-L\left[\frac{\partial f}{\partial \eta_i}-2\kappa_{\eta_i} \left(\frac{\partial^2
    \eta_i}{\partial x^2}+\frac{\partial^2
    \eta_i}{\partial y^2}\right)+\frac{\delta F_{el}}{\delta \eta_i}\right],
\end{equation}
\[i=1,2; x\in [0,127); t\in[0,99).\]
where L is the interfacial kinetic coefficient, M is the concentration dependent mobility, $\kappa_c$ and $\kappa_{\eta}$ are the gradient energy coefficient, and $f(c,\eta)$ is the bulk free energy density which is given as:
\begin{equation} 
\begin{split}
f(c,\eta_1,\eta_2,\eta_3)=&A_1c^2+A_2(1-c)(\eta_1^2+\eta_2^2)\\ &+A_{41}(\eta_1^4+\eta_2^4)+A_{42}(\eta_1^2\eta_2^2)\\&+A_{61}(\eta_1^6+\eta_2^6). 
\end{split}
\end{equation}
The elastic energy is denoted by $F_{el}$ and the variational derivative of elastic free energy is given as:
\begin{equation}
\frac{\delta \mathcal{F}_{el}}{\delta \eta_i}=2\eta_i(r)\left(\sum_{p,i=0}^1B_{pi}(n)\Tilde{\theta}_p(k)\right)_r,    
\end{equation}

\[ \theta_p=\eta_p^2(p=1,2).\]
where, $B_{pi}=C_{stuv}\epsilon_{st}^{(p)}\epsilon_{uv}^{(i)}-\widehat{\sigma}_{st}^{(i)}n_tn_v\omega_{us}(n)\widehat{\sigma}_{uv}^{(p)}$, C is the elastic modulus tensor, $\epsilon$ is the eigen strain, $k$ represents the Fourier wave vector, $n$ is the unit vector in the K-space, $\omega^{-1}_{us}(n)=C_{stuv}n_tn_v$ is the normalized inverse Green tensor, and the elastic stress is given by $\widehat{\sigma}_{st}^{(i)}=C_{stuv}\epsilon_{uv}^{(i)}$.

For this study, two parameters(supersaturated composition and seed) were considered with 5 levels(Table \ref{tab:trainingSet}) i.e, a total of 25($5^2$) combinations.
The training data is represented by (N, T, X, Y)=(5$\times$100$\times$128$\times$128), with N representing the number of instances, T is temporal resolution, and X,Y denoting the spatial resolution along x and y direction, respectively. The input data is the initial condition, while the remaining time frames is output of the model.

\begin{table}[t]
\centering
\begin{tabular}{|c | c | c |}
\hline
Level & \begin{tabular}{@{}c@{}}Supersaturation \\ ($c_0$)\end{tabular}& seed\\
\hline
1&0.20&494\\
2&0.21&1111\\
3&0.22&1482\\
4&0.23&4446\\
5&0.24&7410\\
\hline
\end{tabular}
    \caption{Initial parameters with their levels}
    \label{tab:trainingSet}
\end{table}

\subsection{Methods to compute the derivatives}
The PINO framework can implement different methods of differentiation to compute the equation loss. In this work we have used central difference method(a type of FDM), pseudo spectral method, and Fourier extension. FDM is one of the common numerical techniques that replaces the derivatives by finite difference approximations. To solve the derivatives it divide the domain of the problem into a grid or mesh of discrete points. FDM can handle both periodic and non-periodic boundary conditions.

In the precipitate growth problem, the boundary conditions is periodic in the spacial domain and non-periodic in the time domain. Since, Fourier method works well with periodic problems. We compute the derivative with respect to x and y in the Fourier space and derivative with respect to time using central difference method. This approach is known as  pseudo spectral method. Another approach know as Fourier extension deals with non-periodic boundary conditions by padding zeros in the input to extend the problem domain into a larger and periodic space before performing Fourier differentiation. After computing the derivative in the equation loss with three different numerical derivatives as shown in Table~\ref{tab: derivatives_method}, we found that pseudo-spectral method and Fourier extension improved the Cahn-Hilliard equation loss by twelve orders of magnitude. Moreover, pseudo-spectral method performs better compared to Fourier extension. As a result, we have used pseudo-spectral method to compute the derivatives in this study. We also found that in our PINO model, since the derivatives are calculated in Fourier domain, the additional step of converting fourth derivative of Cahn-Hilliard equation to coupled second order derivative was omitted, thus saving computational costs. This step was performed in PINNs to prevent vanishing gradients~\cite{wight2020solving,mattey2022novel,rezaei2022mixed}.

For $\theta^{\prime}$ precipitate growth in 2D(128$\times$128$\times$100), we found that the inference time of PINO using pseudo-spectral method to calculate the derivatives was 0.332s. Similarly, the inference time was 0.313s for PINO  using Fourier extension to calculate the derivatives. Whereas, the execution time for phase-field model which used pseudo-spectral method to solve the equation was 0.254s. 
\begin{table*}[t]
\centering
\begin{tabular}{|l | c | c | c | c |c|c|c|c|c|c|c|c|}
 \hline
\multicolumn{9}{|c|}{Network with: 64, Network depth:4, Learning rate: 0.0001, Activation function : gelu}\\
\hline
\multicolumn{9}{|c|}{$w_{data(c)}$= 1.0, $w_{data(\eta)}$=1.0,  $w_{PDE(AC)}=1.0$, $w_{PDE(CH)}=0.1$}\\
\hline
Cases & Methods& \begin{tabular}{@{}c@{}}$data_{\eta_1}$ \\ loss\end{tabular}& \begin{tabular}{@{}c@{}}$data_{\eta_2}$ \\ loss\end{tabular}& \begin{tabular}{@{}c@{}}$data_c$  \\ loss\end{tabular}&\begin{tabular}{@{}c@{}}$PDE_{AC(\eta_1)}$ \\ loss\end{tabular}&
\begin{tabular}{@{}c@{}}$PDE_{AC(\eta_2)}$ \\ loss\\\end{tabular} &\begin{tabular}{@{}c@{}}$PDE_{CH(c)}$ \\ loss\\\end{tabular}& \begin{tabular}{@{}c@{}}Total  \\ loss\end{tabular}\\
\hline
1&\begin{tabular}{@{}c@{}}FDM \\ (central)\end{tabular}&9.99$\times 10^{-1}$&9.97$\times 10^{-1}$&9.99$\times 10^{-1}$&7.30$\times 10^{-3}$&6.49$\times 10^{-3}$&8.35$\times 10^{10}$&1.83$\times 10^{10}$\\
2 &\begin{tabular}{@{}c@{}}\textbf{Pseudo- }\\ \textbf{spectral}\end{tabular}&5.33$\times 10^{-3}$&3.53$\times 10^{-3}$&1.91$\times 10^{-2}$&1.13$\times 10^{-3}$&5.40$\times 10^{-4}$&1.05$\times 10^{-1}$&4.03$\times 10^{-2}$\\
3 &\begin{tabular}{@{}c@{}}Fourier \\ extension\end{tabular}&4.58$\times 10^{-3}$&1.01$\times 10^{-2}$&4.60$\times 10^{-2}$&1.56$\times 10^{-3}$&3.27$\times 10^{-3}$&3.31$\times 10^{-1}$&9.87$\times 10^{-2}$\\
\hline
\end{tabular}
\caption{Three different methods to compute the derivatives}
\label{tab: derivatives_method}
\end{table*}

\subsection{Learning the evolution of $\theta^{\prime}$ precipitates using PINO }
Our objective is to compute the solution at each given space and time coordinate from the initial data. Therefore,  the input of the model consist of an initial data of three different fields i.e, $c$, $\eta_1$ and $\eta_2$. To begin with we first determine the optimum number of training data required for training the model. As shown in Table~\ref{tab: training_data}, we see that out of 25 combinations of parameters(Table~\ref{tab:trainingSet}) using 12 different initial conditions give the least training loss for our PINO model.  The algorithm followed for training the model is shown in Algorithm~\ref{alg:algorithm}. 
\begin{table*}
\begin{tabular}{|c| c | c|c |c|c|c|c|c|}
\hline
Cases & \begin{tabular}{@{}c@{}}Training \\ data\end{tabular}& \begin{tabular}{@{}c@{}}$data_{\eta_1}$ \\ loss\end{tabular}& \begin{tabular}{@{}c@{}}$data_{\eta_2}$ \\ loss\end{tabular} &\begin{tabular}{@{}c@{}}$data_c$  \\ loss\end{tabular}&\begin{tabular}{@{}c@{}}$PDE_{\eta_1}$ \\ loss\end{tabular}&
\begin{tabular}{@{}c@{}}$PDE_{\eta_2}$ \\ loss\\\end{tabular} &\begin{tabular}{@{}c@{}}$PDE_{c}$ \\ loss\\\end{tabular} & \begin{tabular}{@{}c@{}}Total  \\ loss\end{tabular}\\
\hline
1 &6&3.26$\times 10^{-3}$&2.58$\times 10^{-3}$&4.70$\times 10^{-2}$&1.02$\times 10^{-3}$&9.70$\times 10^{-4}$&3.44$\times 10^{-1}$&8.92$\times 10^{-2}$\\
2 &8&4.52$\times 10^{-3}$&4.43$\times 10^{-3}$&4.66$\times 10^{-2}$&1.20$\times 10^{-3}$&1.44$\times 10^{-3}$&3.39$\times 10^{-1}$&9.22$\times 10^{-2}$\\
3 &10&4.73$\times 10^{-3}$&3.28$\times 10^{-3}$&4.60$\times 10^{-3}$&1.58$\times 10^{-3}$&1.19$\times 10^{-3}$&3.31$\times 10^{-1}$&8.99$\times 10^{-2}$\\
4 &\textbf{12}&2.95$\times 10^{-3}$&3.07$\times 10^{-3}$&4.51$\times 10^{-2}$&1.05$\times 10^{-3}$&9.90$\times 10^{-4}$&3.33$\times 10^{-1}$&8.65$\times 10^{-2}$\\
5 &14&6.53$\times 10^{-3}$&5.75$\times 10^{-3}$&4.63$\times 10^{-2}$&1.78$\times 10^{-3}$&1.61$\times 10^{-3}$&3.33$\times 10^{-1}$&9.54$\times 10^{-2}$\\
\hline
\end{tabular}
\caption{Determining the optimum number of training data}
\label{tab: training_data}
\end{table*}

\begin{figure}[t]
\centering
\includegraphics[width=\columnwidth]{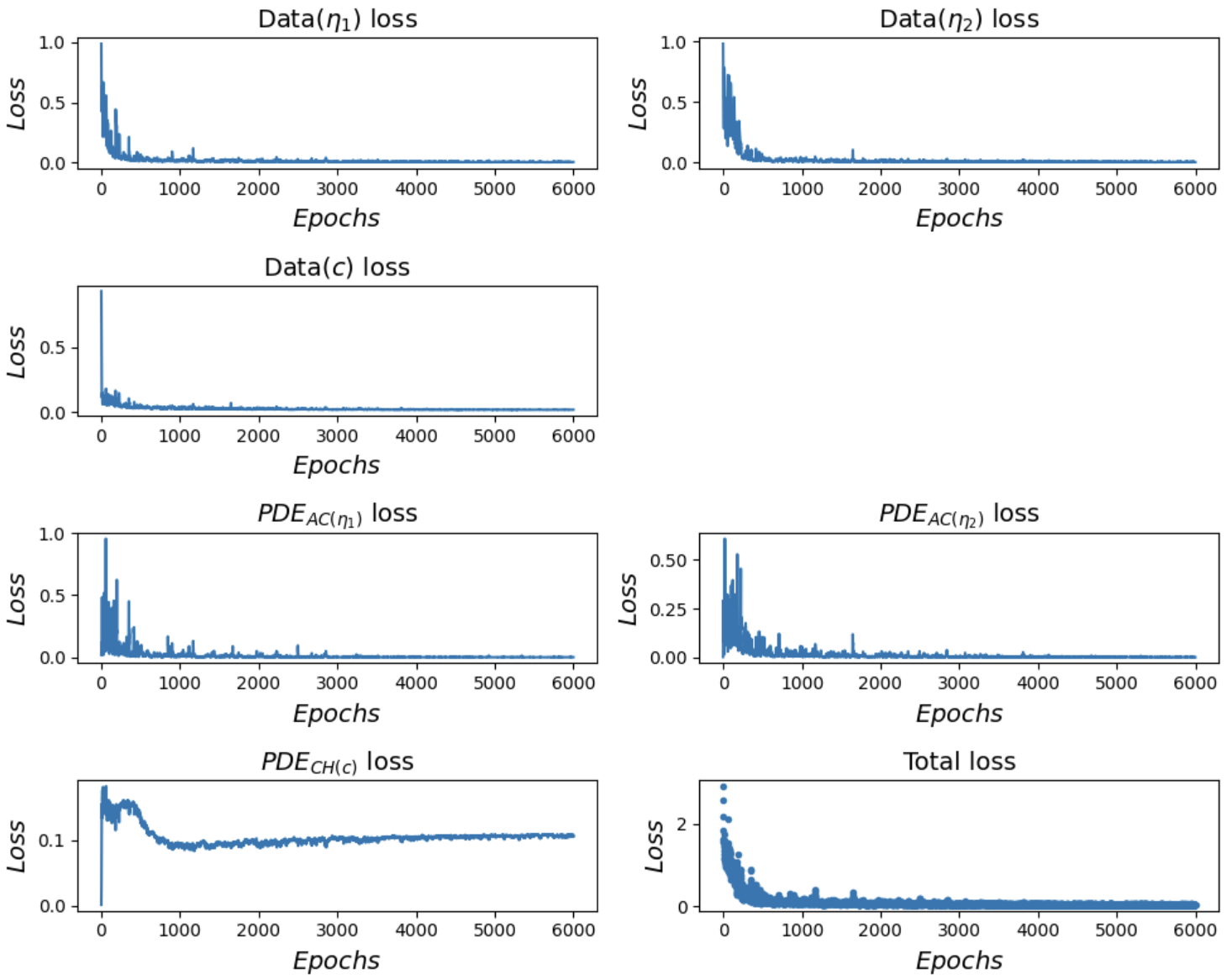}
\caption{Loss vs. epochs} 
\label{fig:loss}
\end{figure}

\begin{figure}[t]
\centering
\includegraphics[width=\columnwidth]{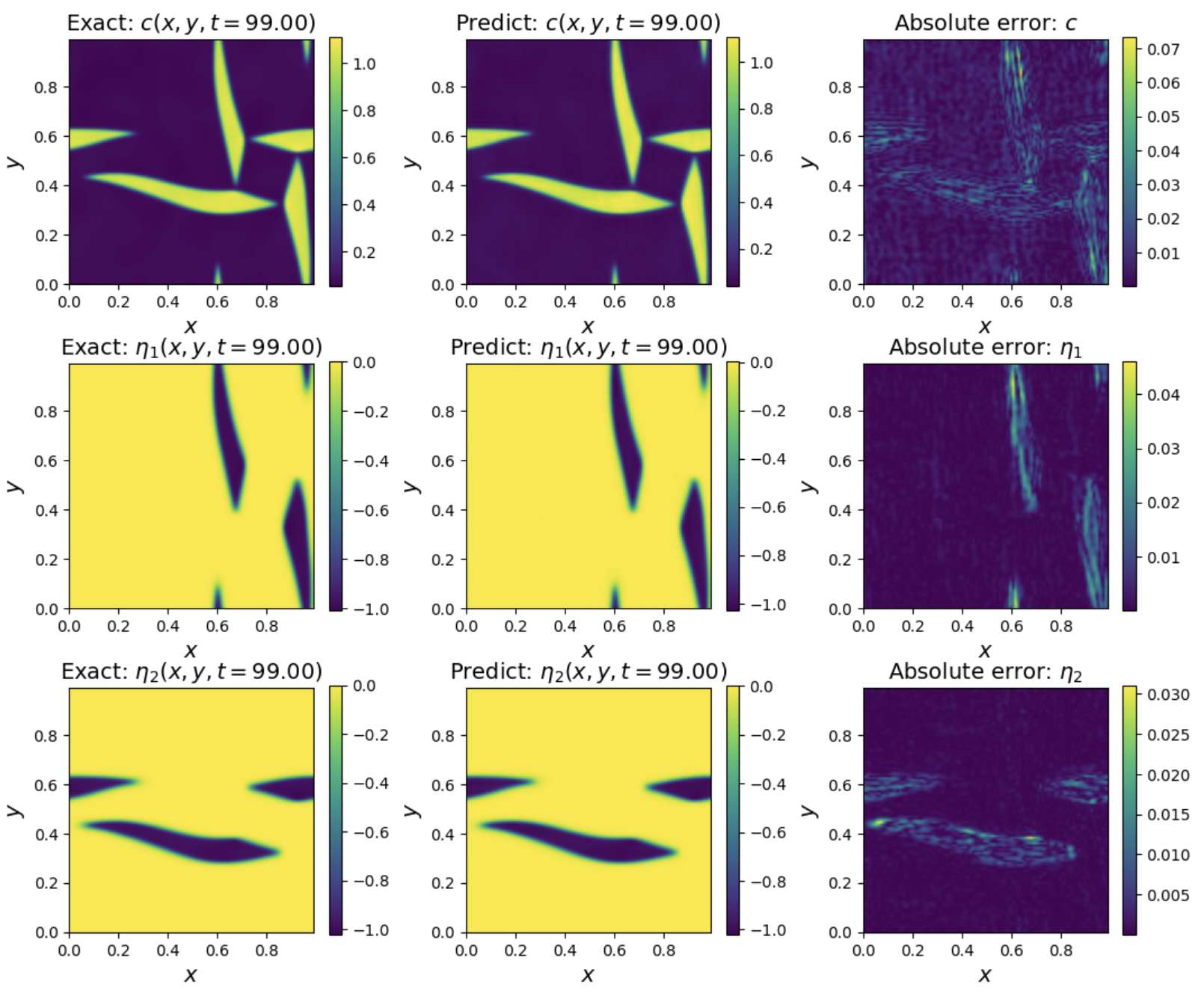}
\caption{Predicting the growth of $\theta^{\prime }$ precipitates for an unseen instance; Supersaturated composition: 0.21,  seed: 1111; Relative L2 error$:c,\eta_1,\eta_2$= 5.14$\times 10^{-2}$, 1.41$\times 10^{-2}$, 1.15$\times 10^{-2}$} 
\label{fig:Predictions1}
\end{figure}

\begin{figure}[t]
\centering
\includegraphics[width=\columnwidth]{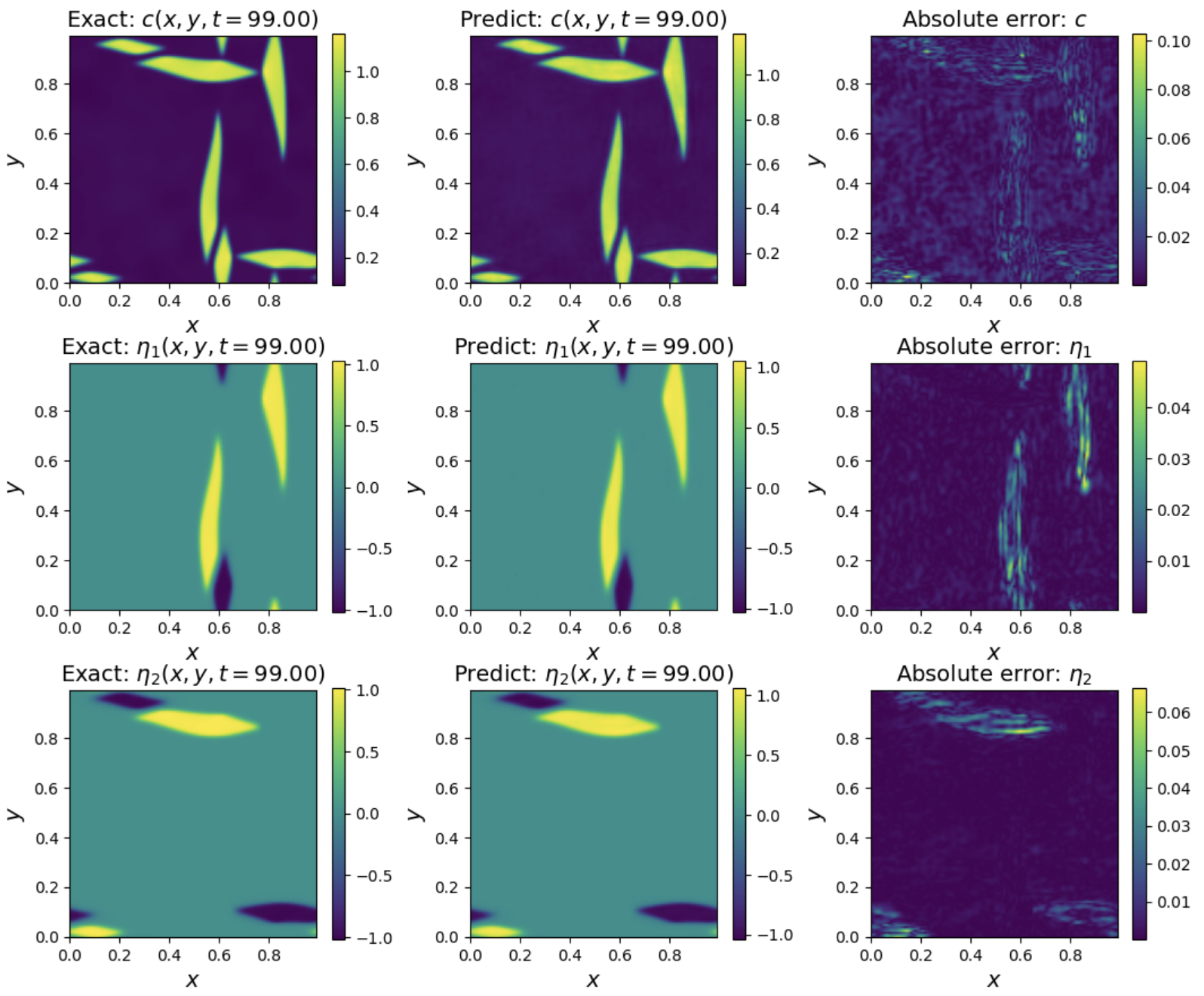}
\caption{Predicting the growth of $\theta^{\prime}$ precipitates for an unseen instance; Supersaturated composition: 0.23, seed: 1482; Relative L2 error$:c,\eta_1,\eta_2$= 5.87$\times 10^{-2}$, 2.21$\times 10^{-2}$, 2.09$\times 10^{-2}$}
\label{fig:Predictions2}
\end{figure}

\begin{algorithm}[!ht]
\caption{Operator Learning: Learn a neural operator $\mathcal{G}_{\theta }$ to approximate $\mathcal{G}^{\dag}$ using both the data loss $\mathcal{L}_{data}$ and the PDE loss $\mathcal{L}_{pde}$.}
\label{alg:algorithm}
\textbf{Input}: input output function pairs $\{a_j,u_j\}_{j=1}^K$\\
\textbf{Output}: Neural operator \[\mathcal{G}: \mathcal{A}\rightarrow \mathcal{U}\]
\begin{algorithmic}[1] 
\FOR{i = 0,1,2,$\dots$}
      \FOR{j =1 to K}
      \STATE Lift the input $a_j\in \mathcal{A}$ to a higher dimensional representation by the local transformation P
        \[\upsilon_n(x,y,t)=P(a(x,y,t))\]
        \FOR {n = 1 to N no. of Fourier layers\dots}
        \STATE Update the Fourier layer \\
        \
        \[\upsilon_n\longrightarrow \upsilon_{n+1}\]
        
        \STATE Project the output from the last fourier layer to  original dimension
        \[u(x,y,t)=Q(\upsilon_N(x,y,t))\]
         \ENDFOR
        \STATE Compute
  $\mathcal{L}_{data}$  and $\mathcal{L}_{pde}$. 

  \STATE Update neural operator $\mathcal{G}$
      \ENDFOR
    \ENDFOR
\end{algorithmic}
\end{algorithm}
Similarly, by trial and error approach we optimize the architecture of PINO. The Fourier layer, modes, and width of the PINO model were chosen as  4, 28, and 64 respectively. We observed convergence in the loss when the number of epochs was about 6000 as shown in Fig.~\ref{fig:loss}.
The model's hyper-parameters, which include the learning rate and gamma value, were selected to be 0.9 and 0.001, respectively. The weights of loss terms are taken as $w_{data(\eta)}=1$, $w_{pde(AC)}=1$, $w_{data(c)}=1$,$w_{pde(CH)}=0.1$. The accuracy of the model was assessed using L2 norm of the relative error. A comparison of the predictions from our PINO model with that from high-fidelity phase-field simulations at time step 99 for two different unseen initial condition is shown in Fig.~\ref{fig:Predictions1} and Fig.~\ref{fig:Predictions2}. This shows that the trained PINO model has the capability to predict the growth of $\theta^{\prime}$ precipitates with good accuracy for those instances or initial conditions which were not seen during training.

\section{Conclusion}
We have demonstrated the ability of PINOs to accurately learn the physics of two second-order and one fourth-order coupled PDEs with high accuracy. PINO was able to handle higher order PDEs with good accuracy since it calculated the derivatives in the Fourier domain. We have also demonstrated the capability of PINOs to generalize to other instances without re-training. Although we have shown the efficiency of PINO model on precipitation growth problem, it can be used for predicting other complex microstructural evolution problems such as grain growth and coarsening, solidification, and crack propagation.

For a 2D problem with a domain size of 128$\times$128$\times$100, we did not see an improvement in computational time. This could be because numerical approaches for smaller spatial domain require less computing time. We may need to go to larger spacial domain to see the computational advantage of PINO over numerical approaches.

\section{Future work}
Most of the studies on $\theta^{\prime}$ precipitation are limited to 2D simulation due to the enormous computational power required for 3D phase-field simulations. Therefore, our goal is to extend the current 2D PINO model to 3D PINO model to predict the evolution of $\theta^{\prime}$ precipitates in Al-Cu binary alloys. We will  explore the zero-shot resolution capability(prediction of unseen instances at higher resolution compared to those used in training) of PINOs to reduce the computational cost. In this study we have shown that PINOs can learn the physics of coupled higher order PDEs with high accuracy. We further want to perform temporal extrapolation tests as well as determine the prediction capability of PINO to unseen initial conditions in extrapolation settings.


\bibliography{arxiv_Ggangmei}

\begin{thebibliography}{17}
\providecommand{\natexlab}[1]{#1}

\bibitem[{Allen and Cahn(1979)}]{allen1979microscopic}
Allen, S.~M.; and Cahn, J.~W. 1979.
\newblock A microscopic theory for antiphase boundary motion and its application to antiphase domain coarsening.
\newblock \emph{Acta metallurgica}, 27(6): 1085--1095.

\bibitem[{Barrett, Blowey, and Garcke(1999)}]{barrett1999finite}
Barrett, J.~W.; Blowey, J.~F.; and Garcke, H. 1999.
\newblock Finite element approximation of the Cahn--Hilliard equation with degenerate mobility.
\newblock \emph{SIAM Journal on Numerical Analysis}, 37(1): 286--318.

\bibitem[{Cahn(1961)}]{cahn1961spinodal}
Cahn, J.~W. 1961.
\newblock On spinodal decomposition.
\newblock \emph{Acta metallurgica}, 9(9): 795--801.

\bibitem[{Chen et~al.(2024)Chen, Lucarini, Ma, Chen, and Cui}]{chen4761824pf}
Chen, N.; Lucarini, S.; Ma, R.; Chen, A.; and Cui, C. 2024.
\newblock Pf-Pinns: Physics-Informed Neural Networks for Solving Coupled Allen-Cahn and Cahn-Hilliard Phase Field Equations.
\newblock \emph{Available at SSRN 4761824}.

\bibitem[{Kumar~Makineni et~al.(2017)Kumar~Makineni, Sugathan, Meher, Banerjee, Bhattacharya, Kumar, and Chattopadhyay}]{kumar2017enhancing}
Kumar~Makineni, S.; Sugathan, S.; Meher, S.; Banerjee, R.; Bhattacharya, S.; Kumar, S.; and Chattopadhyay, K. 2017.
\newblock Enhancing elevated temperature strength of copper containing aluminium alloys by forming L12 Al3Zr precipitates and nucleating $\theta^{\prime\prime}$ precipitates on them.
\newblock \emph{Scientific reports}, 7(1): 1--9.

\bibitem[{Li et~al.(2020)Li, Kovachki, Azizzadenesheli, Liu, Bhattacharya, Stuart, and Anandkumar}]{li2020fourier}
Li, Z.; Kovachki, N.; Azizzadenesheli, K.; Liu, B.; Bhattacharya, K.; Stuart, A.; and Anandkumar, A. 2020.
\newblock Fourier neural operator for parametric partial differential equations.
\newblock \emph{arXiv preprint arXiv:2010.08895}.

\bibitem[{Li et~al.(2021)Li, Zheng, Kovachki, Jin, Chen, Liu, Azizzadenesheli, and Anandkumar}]{li2021physics}
Li, Z.; Zheng, H.; Kovachki, N.; Jin, D.; Chen, H.; Liu, B.; Azizzadenesheli, K.; and Anandkumar, A. 2021.
\newblock Physics-informed neural operator for learning partial differential equations.
\newblock \emph{ACM/JMS Journal of Data Science}.

\bibitem[{Liu and Shen(2003)}]{liu2003phase}
Liu, C.; and Shen, J. 2003.
\newblock A phase field model for the mixture of two incompressible fluids and its approximation by a Fourier-spectral method.
\newblock \emph{Physica D: Nonlinear Phenomena}, 179(3-4): 211--228.

\bibitem[{Lu et~al.(2021)Lu, Jin, Pang, Zhang, and Karniadakis}]{lu2021learning}
Lu, L.; Jin, P.; Pang, G.; Zhang, Z.; and Karniadakis, G.~E. 2021.
\newblock Learning nonlinear operators via DeepONet based on the universal approximation theorem of operators.
\newblock \emph{Nature machine intelligence}, 3(3): 218--229.

\bibitem[{Mattey and Ghosh(2022)}]{mattey2022novel}
Mattey, R.; and Ghosh, S. 2022.
\newblock A novel sequential method to train physics informed neural networks for Allen Cahn and Cahn Hilliard equations.
\newblock \emph{Computer Methods in Applied Mechanics and Engineering}, 390: 114474.

\bibitem[{Paszke et~al.(2019)Paszke, Gross, Massa, Lerer, Bradbury, Chanan, Killeen, Lin, Gimelshein, Antiga et~al.}]{paszke2019pytorch}
Paszke, A.; Gross, S.; Massa, F.; Lerer, A.; Bradbury, J.; Chanan, G.; Killeen, T.; Lin, Z.; Gimelshein, N.; Antiga, L.; et~al. 2019.
\newblock Pytorch: An imperative style, high-performance deep learning library.
\newblock \emph{Advances in neural information processing systems}, 32.

\bibitem[{Raissi, Perdikaris, and Karniadakis(2019)}]{raissi2019physics}
Raissi, M.; Perdikaris, P.; and Karniadakis, G.~E. 2019.
\newblock Physics-informed neural networks: A deep learning framework for solving forward and inverse problems involving nonlinear partial differential equations.
\newblock \emph{Journal of Computational physics}, 378: 686--707.

\bibitem[{Rezaei et~al.(2022)Rezaei, Harandi, Moeineddin, Xu, and Reese}]{rezaei2022mixed}
Rezaei, S.; Harandi, A.; Moeineddin, A.; Xu, B.-X.; and Reese, S. 2022.
\newblock A mixed formulation for physics-informed neural networks as a potential solver for engineering problems in heterogeneous domains: Comparison with finite element method.
\newblock \emph{Computer Methods in Applied Mechanics and Engineering}, 401: 115616.

\bibitem[{Rosofsky, Al~Majed, and Huerta(2023)}]{rosofsky2023applications}
Rosofsky, S.~G.; Al~Majed, H.; and Huerta, E. 2023.
\newblock Applications of physics informed neural operators.
\newblock \emph{Machine Learning: Science and Technology}, 4(2): 025022.

\bibitem[{Sun(1995)}]{sun1995second}
Sun, Z.~Z. 1995.
\newblock A second-order accurate linearized difference scheme for the two-dimensional Cahn-Hilliard equation.
\newblock \emph{Mathematics of Computation}, 64(212): 1463--1471.

\bibitem[{Wang, Wang, and Perdikaris(2021)}]{wang2021learning}
Wang, S.; Wang, H.; and Perdikaris, P. 2021.
\newblock Learning the solution operator of parametric partial differential equations with physics-informed DeepONets.
\newblock \emph{Science advances}, 7(40): eabi8605.

\bibitem[{Wight and Zhao(2020)}]{wight2020solving}
Wight, C.~L.; and Zhao, J. 2020.
\newblock Solving Allen-Cahn and Cahn-Hilliard equations using the adaptive physics informed neural networks.
\newblock \emph{arXiv preprint arXiv:2007.04542}.

\end{thebibliography}

\end{document}